\documentclass[aps,prc,preprint,groupedaddress,amsmath]{revtex4-1} 
\usepackage[dvipdfmx]{graphicx}
\usepackage{braket}
\usepackage{bm}
\usepackage{longtable}

\usepackage{amsmath}
\begin{document}

\title{Hoyle-analogue state in $\bf{^{13}}$C 
  \\studied with Antisymmetrized Molecular Dynamics}

\author{Y. Chiba}
\affiliation{Department of Physics, Hokkaido University, 060-0810 Sapporo, Japan}
\author{M. Kimura}
\affiliation{Department of Physics, Hokkaido University, 060-0810 Sapporo, Japan}
\affiliation{Nuclear Reaction Data Centre, Faculty of Science, Hokkaido University, Sapporo
060-0810, Japan} 

\date{\today}

\begin{abstract}
 The cluster states in $^{13}{\rm C}$ are investigated by antisymmetrized molecular dynamics. By
 investigating the spectroscopic 
 factors, the cluster configurations of the excited states are discussed. It is found that the
 $1/2^+_2$ state is dominantly composed of the $^{12}{\rm C}(0^+_2)\otimes s_{1/2}$ configuration
 and can be regarded as a Hoyle analogue state. On the other hand, the $p$-wave states ($3/2^-$
 and $1/2^-$) do not have such structure, because of the coupling with other configurations. The
 isoscalar monopole and dipole transition strengths from the ground to the excited states  are
 also studied. It is shown that the excited $1/2^-$ states have strong isoscalar monopole
 transition strengths consistent with the observation. On the other hand, the excited $1/2^+$
 states unexpectedly have weak isoscalar dipole  transitions except for the $1/2^+_1$ state.
 It is discussed that the suppression of the dipole transition is attributed to the property of 
 the dipole operator.
\end{abstract}

\pacs{Valid PACS appear here}% PACS, the Physics and Astronomy
\maketitle

\section{Introduction}
In these decades, the Hoyle state of $^{12}$C
\cite{UegakiPTP57,KamimuraNPA351,EnyoPRL81,THSR,FunakiPRC,NeffPRL98} attracts much interests as a
possible bosonic condensate. A natural extension of the discussion is the search for the analogue
states in heavier $4n$ nuclei such as $^{16}$O and $^{20}$Ne.  Recently, a possible
candidate in $^{16}$O \cite{FunakiPRC80,O1,O2,O3} is being intensively discussed, and a
theoretical study  
\cite{YamadaPRC69} predicted the existence of the $\alpha$ particle condensate up to approximately
10 $\alpha$ system $^{40}$Ca. 

Another direction of the research is the study of $N\neq Z$ nuclei in which nucleon particles or
holes can be injected into the $\alpha$ particle condensate as an impurity. In the case of
$^{11}$B which has a proton hole coupled to $^{12}$C, the theoretical studies based on
antisymmetrized molecular dynamics (AMD) \cite{EnyoPRC75, SuharaB} pointed out that the 
$3/2^-_3$ state located just below the $^7\mathrm{Li}+\alpha$ threshold has pronounced $2\alpha+t$
clustering with large radius. Hence, the state was suggested as  a candidate of the Hoyle analogue
state. More recently, T. Yamada {\it et al.} performed the orthogonality condition model (OCM)
calculation  \cite{YamadaPRC82} and predicted the $1/2^+_2$ state as another candidate of Hoyle
analogue state in which all of $2\alpha$ and triton particles occupy the $s$-wave state.

Several discussions have also been made for $^{13}$C which has an extra neutron.  T. Yamada
 {\it et al.} \cite{YamadaMPLA21} discussed the possible reduction of spin-orbit splitting in the
 Hoyle analogue states. Namely, they suggested that the spin-orbit splitting between the $p_{1/2}$ 
and $p_{3/2}$ coupled to the Hoyle state ($1/2^-$ and $3/2^-$ states) will be reduced, because the
splitting is dependent on the first derivative of the density distribution and the Hoyle state has
dilute density profile. In addition to this, T. Yamada {\it et al.} performed the OCM calculation
\cite{YamadaPRC92} and predicted the Hoyle-analogue $1/2^+$ state in which all of three $\alpha$
particles and a neutron occupy the same $s$-wave state, that is quite similar to the discussion
made for  $^{11}$B. Thus, the $3/2^-$, $1/2^-$ and $1/2^+$ states in $^{13}$C are of particular
interest  and importance for the understanding of  the Hoyle analogue state in $N \neq Z$ nuclei. 

Up to now, the existence of the Hoyle-analogue $3/2^-$, $1/2^-$ and $1/2^+$ states in $^{13}$C is
still ambiguous, because the information is not enough to identify them. Therefore, in this work,
we conduct the AMD calculation to supply further theoretical information.  We investigate the
spectroscopic factors ($S$-factors) in $^{12}\mathrm{C}+n$ and $^{9}\mathrm{Be}+\alpha$ channels
to identify the Hoyle analogue states. Furthermore,  we focus on the isoscalar dipole (IS1)
transition strength as well as the isoscalar monopole (IS0) transition strength, which are known 
to be enhanced for the cluster states
\cite{KawabataB,EnyoB,YamadaPTP120,YamadaPRC85,ChibaPRC93,KimuraEPJA}. We expect that they are useful to 
identify the Hoyle analogue states in $^{13}{\rm C}$. 

This paper is organized as follows: First, we explain theoretical framework of AMD and how to
calculate $S$-factors of the $^{12}\mathrm{C}+n$ and $^{9}\mathrm{Be}+\alpha$ channels.  Second,
we present our numerical calculation results and compare it to experimental data.  We analyze
nuclear structure of $3/2^-$, $1/2^-$ and $1/2^+$ states in detail using $S$-factors and identify
the Hoyle-analogue states. We also discuss on the IS0 and IS1 transition strengths  to supply
theoretical information for forthcoming experiments.  Finally, we summarize our paper.  

\section{AMD Framework}
\subsection{Hamiltonian and model wave function}
The Hamiltonian employed in this work is 
\begin{align}
  \hat{H} = \sum_{i=1}^A \hat{t}_i - \hat{t}_{c.m.} + \sum_{i<j} \hat{v}_{NN} + \sum_{i<j} \hat{v}_{Coul},
\end{align}
where $\hat{t}_i$ is the $i$-th nucleon kinetic energy and $\hat{v}_{NN}$ and
$\hat{v}_{Coul}$ are the Gogny D1S nucleon-nucleon interaction \cite{BergerCPC63} and Coulomb
interaction, respectively. The center-of-mass kinetic energy
$\hat{t}_{c.m.}$ is subtracted from the Hamiltonian.

The intrinsic AMD wave function is a Slater determinant of nucleon Gaussian wave packets
\cite{PTPS192,AMDCRP,KimuraPRC69}, 
\begin{align}
  \Phi_{AMD} &= \mathcal{A}\left\{ \varphi_1\varphi_2 \dots\varphi_A \right\},
  \\
  \varphi_i &= \phi_i \otimes \chi_i \otimes \xi_i,\\
  \phi_i &= \left( \frac{\pi^3}{8|\bm{\nu}|} \right)^{-\frac{1}{4}}
    \exp{\left[ -\sum_{\sigma=xyz} \nu_\sigma \left( r_{i\sigma} - \frac{Z_{i\sigma}}{\sqrt{\nu_\sigma}} \right) \right] },
  \\
  \chi_i &= \alpha_i \ket{\uparrow} + \beta_i \ket{\downarrow},\quad
  \xi_i = \ket{p} \text{ or } \ket{n}.
\end{align}
It is noted that the  center-of-mass wave function $\Phi_{c.m.}$  is analytically separable from the intrinsic 
wave function,
\begin{align}
  \Phi_{AMD} &= \Phi_{int} \Phi_{c.m.},\\
  \Phi_{c.m.} &= \left( \frac{\pi^3}{8A^3|\bm{\nu}|} \right)^{-\frac{1}{4}}
  \exp{\left[ - A \sum_{\sigma=xyz} \nu_{\sigma} R^2_{\sigma} \right] }.
\end{align}
Here, $\Phi_{int}$ is the internal wave function, and we assume 
that the relation $\sum_{i}
\bm{Z}_i = \bm{0}$ holds. Therefore, the AMD framework is completely free from spurious
motion. 
This is an important advantage when we calculate the IS1 transition
strengths.
The parameters of AMD wave function $\bm{\nu}$, $\bm{Z}_i$,
$\alpha_i$ and $\beta_i$ are determined so as to
minimize the energy after parity-projection,
\begin{align}
  &\Phi^\pi = \frac{1+\pi\hat{P}_x}{2} \Phi_{int},\ \pi = \pm,
  \\ &E^{\pi} = \frac{\braket{\Phi^{\pi}|\hat{H}|\Phi^{\pi}}}{\braket{\Phi^\pi|\Phi^\pi}}.
\end{align}

To describe the various states of $^{13}$C, we impose the constraint on the expectation values of harmonic oscillator quanta $N$,
$\lambda$ and $\mu$ which are defined by using the harmonic
oscillator quanta in
Cartesian coordinate $N_x,N_y$ and $N_z$,
\begin{align}
  N = N_x + N_y + N_z,\quad \lambda = N_z - N_y,\quad \mu= N_y - N_x.
\end{align}
Here, we assume the relation $N_x \leq N_y \leq N_z$. Roughly speaking, the excitation of
system is expressed by $N$, and $\lambda$
and $\mu$ indicate the asymmetries around the longest and shortest deformed axis.
The detail of this constraint is described in Ref.~\cite{ChibaPRC91}.

Compared with 
the constraint on the quadrupole deformation parameters
($\beta\gamma$-constraint), which is often
used in mean-field and AMD calculations, the
constraint on $N$, $\lambda$ and $\mu$ is appropriate
for the description
of the highly excited states. The $\beta\gamma$-constraint is
useful to describe the low-lying quadrupole collectivities but it
often fails to describe highly excited states. On the other hand, 
the constraint on $N$, $\lambda$ and $\mu$ is
capable to describe the highly excited states 
with many-particle many-hole configurations.
In this study, the possible combinations of values for $N$, $\lambda$ and $\mu$ up to
$N=18$ ($9\hbar\omega$ excitation) are adopted as 
the constraint. We
denote thus-obtained basis wave function as $\Phi^\pi(N\lambda\mu)$.

After energy variation, the basis wave functions are projected to
angular-momentum eigenstates and superposed  to obtain excitation spectra
and eigen wave functions (generator coordinate method (GCM)).
\begin{align}
  \Phi^{J\pi}_{MK}(N_i\lambda_i\mu_i) = \mathcal{N}_K^{-\frac{1}{2}}\hat{P}^J_{MK} \Phi^\pi(N_i\lambda_i\mu_i),\\
  \mathcal{N}_K = \braket{\Phi^{J\pi}_{MK}(N\lambda\mu)|\Phi^{J\pi}_{MK}(N\lambda\mu)},\\
  \Psi^{J\pi}_n = \sum_{Ki} g^{J\pi}_{Kin} \Phi^{J\pi}_{MK}(N_i\lambda_i\mu_i),
\end{align}
where $\hat{P}^{J}_{MK}$ is the angular momentum projection
operator.
The coefficients $g^{J\pi}_{Kin}$ is determined by diagonalizing the Hamiltonian,
\begin{align}
  &\sum_{i'K'} H^{J\pi}_{iKi'K'} g^{J\pi}_{i'K'n} = E^{J\pi}_n \sum_{i'K'} N^{J\pi}_{iKi'K'}g^{J\pi}_{i'K'n},\\
  &H^{J\pi}_{iKi'K'} = \braket{\Phi^{J\pi}_{MK}(N_i\lambda_i\mu_i)|\hat{H}|\Phi^{J\pi}_{MK'}(N_{i'}\lambda_{i'}\mu_{i'})},\\
  &N^{J\pi}_{iKi'K'} = \braket{\Phi^{J\pi}_{MK}(N_i\lambda_i\mu_i)|\Phi^{J\pi}_{MK'}(N_{i'}\lambda_{i'}\mu_{i'})}.
\end{align}

\subsection{Reduced width amplitudes and spectroscopic factors}
To search for Hoyle-analogue states, we calculate the reduced width
amplitudes and $S$-factors in the $^{12}$C+$n$ and $^9$Be+$\alpha$ channels. The
reduced width amplitude in the $^{12}$C+$n$ channel is defined as 
\begin{align}
  y^{J\pi n}_{j_{\mathrm{C}}\pi_{\mathrm{C}}jl}(a) = \sqrt{13} 
  \Braket{\frac{\delta(r-a)}{r^2}\left[ \Phi^{j_{\mathrm{C}}\pi_{\mathrm{C}}}_{\mathrm{C}}  \left[ Y_{l}(\hat{r})\chi_{1/2} \right]_{j} \right ]_{J}
  |\Psi^{J\pi}_{n}}
\end{align}
where $\Phi^{j_{\mathrm{C}}\pi_{\mathrm{C}}}_{\mathrm{C}}$ is the wave function of $^{12}$C, and
$\chi_{1/2}$ is the spin-isospin wave function of neutron.  $j_{\mathrm{C}}$
and $\pi_{\mathrm{C}}$ are angular momentum and parity of $^{12}$C, and $j$ and $l$ are
total and orbital angular momenta of neutron.
In a same manner, the reduced width amplitude in the
$^{9}$Be+$\alpha$ channel is defined as
\begin{align}
  y^{J\pi n}_{j_{\mathrm{Be}}\pi_\mathrm{Be} l}(a) = \sqrt{\frac{13!}{9!4!}}
  \Braket{ \frac{\delta(r-a)}{r^2} \Phi_\alpha  \left[ \Phi^{j_\mathrm{Be}\pi_\mathrm{Be}}_{\mathrm{Be}}Y_{l}(\hat{r}) \right]_{J} | \Psi^{J\pi}_n}
\end{align}
where $\Phi^{j_{\mathrm{Be}}\pi_{\mathrm{Be}}}_{\mathrm{Be}}$ is  the wave function of $^{9}$Be with angular momentum $j_\mathrm{Be}$ and parity $\pi_\mathrm{Be}$.
$\Phi_{\alpha}$ is the wave function of the ground state of $\alpha$ cluster.
The $S$-factors of the $^{12}$C+$n$ and $^{9}$Be+$\alpha$ channels are defined as
the integral of the reduced width amplitudes,
\begin{align}
  S^{J\pi n}_{j_{\mathrm{C}}\pi_\mathrm{C}jl} &= \int^{\infty}_{0} da \ \left| ay^{J\pi n}_{j_{\mathrm{C}}\pi_{\mathrm{C}}jl}(a) \right|^2,\\
  S^{J\pi n}_{j_{\mathrm{Be}}\pi_{\mathrm{Be}}l} &= \int^{\infty}_{0} da \ \left| ay^{J\pi n}_{j_{\mathrm{Be}}\pi_{\mathrm{Be}}l}(a)\right|^2.
\end{align}
%The $S$-factor is the measure of subsystem components included in the excited states of $^{13}$C.

To evaluate the reduced width amplitudes, we use Laplace expansion method
proposed in Ref.~\cite{ChibaPTEP}, which can treat the deformed
and the excited cluster wave functions without any approximations. In this study,
the  $\alpha$ cluster is described by the $(0s)^4$
configuration with oscillator parameter $\nu = {m\omega}/{(2\hbar)} = 0.25$ fm$^{-1}$ while
the wave functions of $^9$Be and $^{12}$C are obtained by AMD+GCM calculations.

\section{Results and discussions}  \label{sec:3}

\subsection{Excitation spectra}
\begin{figure*}
  \includegraphics[width=0.8\hsize]{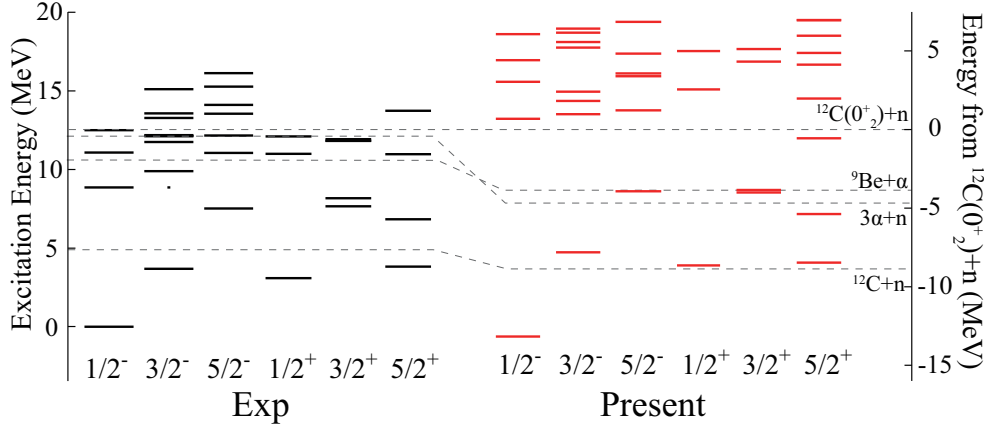}
  \caption{(Color online) The calculated and observed \cite{NDA} excitation spectra of the 
 $1/2^\pm$, $3/2^\pm$ and $5/2^\pm$ states of $^{13}{\rm C}$.} 
  \label{fig:spectra}
\end{figure*}

We performed the energy variation under constraint on
the harmonic oscillator quanta $N$, $\lambda$ and $\mu$.
Using the basis wave functions generated by the energy variation, we 
performed the GCM calculation and obtained excitation spectra of $^{13}$C.
The observed data \cite{NDA} and calculated
excitation spectra up to $E_x= 20$ MeV with $J^\pi\leq5/2^\pm$ are shown in Fig.~\ref{fig:spectra}. 

The calculated yrast states reasonably agree with the observed spectra. 
On the other hand, the energies of the non-yrast states above
10 MeV are overestimated.
For example, the $3/2^-_2$ state observed at 9.9 MeV is 
located at 13.2 MeV in the present calculation. 
As shown later,
many of the excited states located above 10 MeV 
have cluster structure. Thus, we can say that the present calculation
overestimates the energies of the cluster states.
This is mainly due to the limitation of our model space.
The restriction
up to $N=18$ configuration may not be sufficient
to describe the relative motion of clusters.

\subsection{Structure of $1/2^-$ states}
\begin{figure}
  \includegraphics[width=0.7\hsize]{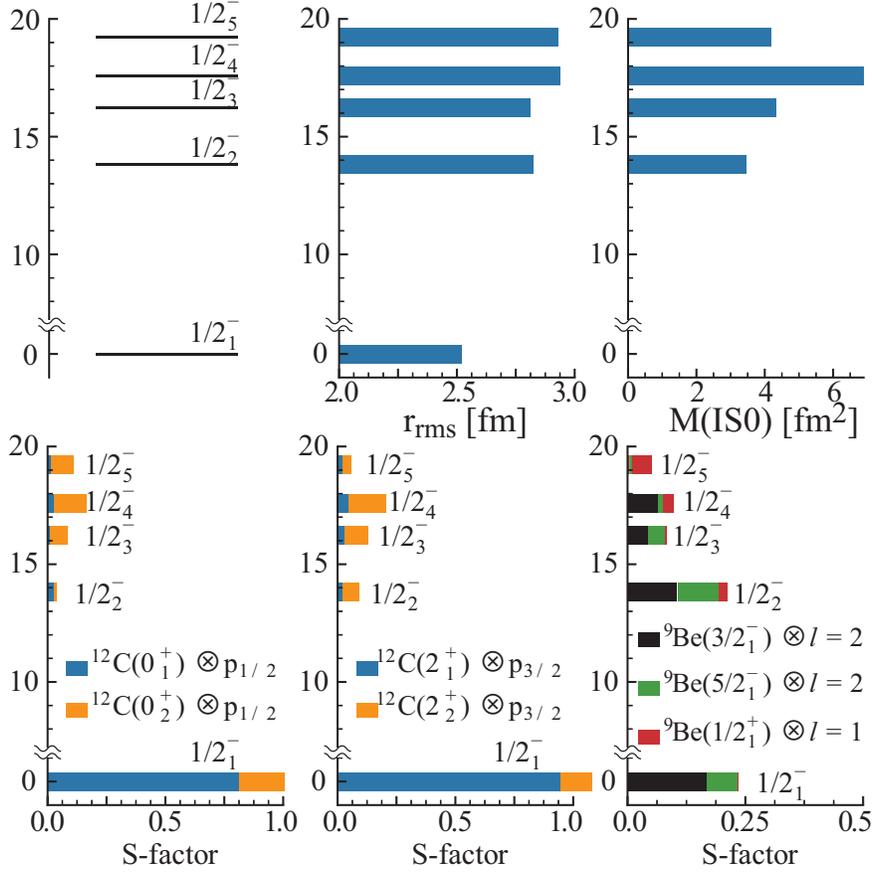}
  \caption{(Color online) The excitation energies and the properties of the $1/2^-$ states
 obtained in the present work. In the upper panels, the calculated excitation spectra, matter rms
 radii $r_{rms}$,  IS0 transition matrix from the ground state $M(IS0)$ are shown from left to
 right. The calculated $S$-factors in the $^{12}\mathrm{C}+n$ and $^9\mathrm{Be}+\alpha$
 channels are presented in the lower panels.   In the last panel, $l$ denotes the orbital 
 angular momentum between $\alpha$ and $^{9}{\rm B}$ clusters. } 
 \label{fig:sfactor1-}
\end{figure}
\begin{figure}
  \includegraphics[width=0.8\hsize]{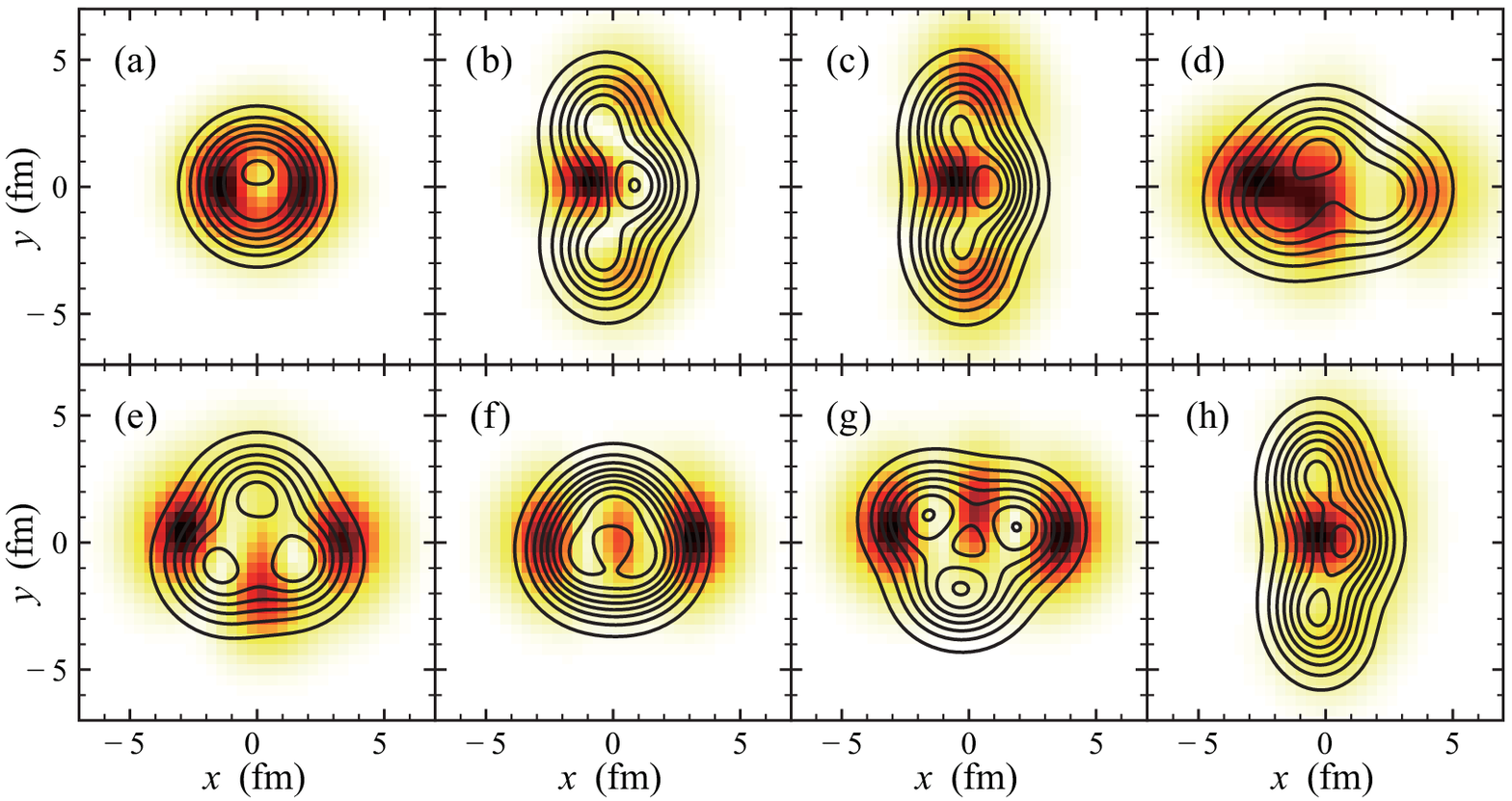}
  \caption{(Color online) 
    Intrinsic matter and valence neutron density distributions of the basis
  wave functions obtained with the variation under the constraint on the H.O.
quanta on $z=0$ plane. Contour plot indicates the matter density distribution while the color plot
indicates the valence neutron density distribution.}
  \label{fig:den}
\end{figure}

In this section, we discuss the structure of the $1/2^-$ states,
which are the candidates of the Hoyle-analogue state having $^{12}\mathrm{C}(0^+_2)\otimes p_{1/2}$ configuration.
The calculated root-mean-square (rms) radii,
$S$-factors in the $^{12}\mathrm{C}+n$ and $^9\mathrm{Be}+\alpha$ channels
and the IS0 transition matrix from the ground state $M(IS0)$ are shown in Fig.~\ref{fig:sfactor1-}. 
The transition matrix is defined as,
\begin{align}
  &\mathcal{M}^{IS0} = \sum_{i=1}^{A} \left(\bm{r}_i-\bm{r}_{c.m.}\right)^2,\ \bm{r}_{c.m.} = \frac{1}{A}\sum_i \bm{r}_i,\\
  &M(IS0) = \left|\Braket{\Psi^{1/2-}_k|\mathcal{M}^{IS0}|\Psi^{1/2-}_{1}}\right|.
\end{align}

The ground state ($1/2^-_1$ state) has
the compact shell structure with the radius of $2.52$ fm.
This state has large overlap (0.94) with the basis wave function
obtained by the energy variation with the constraint $(N,\lambda,\mu)=(9,0,3)$ whose intrinsic
density distribution is shown in Fig.~\ref{fig:den} (a).
The shell-model like character of the ground state can be
confirmed by the large $S$-factors in the 
$^{12}\mathrm{C}(0^+_1)\otimes p_{1/2}$
and $^{12}\mathrm{C}(2^+_1)\otimes p_{3/2}$ channels, which 
are 0.81 and 0.94 respectively.
It is noted that $^{12}\mathrm{C}(0^+_1)\otimes p_{1/2}$ and $^{12}\mathrm{C}(2^+_1)\otimes p_{3/2}$
channels identically corresponds to $(0s)^4(0p_{3/2})^8(0p_{1/2})^1$ configuration
if the $^{12}\mathrm{C}(0^+_1)$ and $^{12}\mathrm{C}(2^+_1)$ have the
$(0s)^4(0p_{3/2})^8$ and $(0s)^4(0p_{3/2})^7(0p_{1/2})^1$ configurations, respectively.

\begin{figure}
  \includegraphics[width=0.5\hsize]{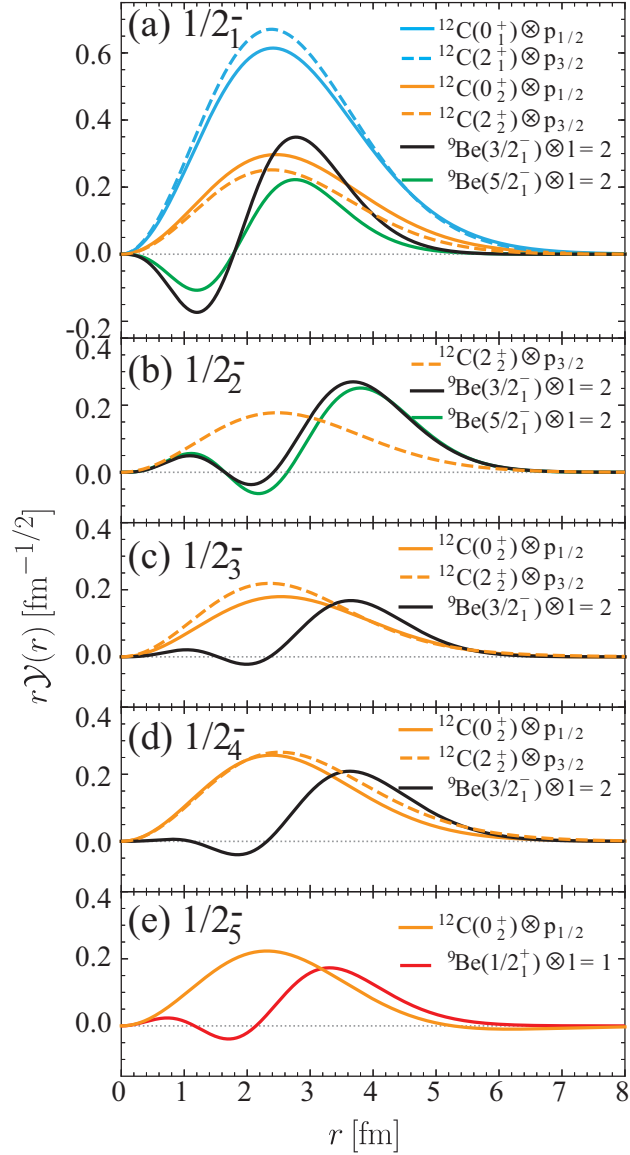}
  \caption{(Color online) The calculated RWAs of the $1/2^-$ states in the $^{12}\mathrm{C}+n$ 
    and $^9\mathrm{Be}+\alpha$ channels. The RWAs which yields the larger $S$ than
    0.04 are displayed.}
 \label{fig:rwa1-}
\end{figure}

While the ground state has the compact shell structure, the excited $1/2^-$
states have the larger rms radii than 2.70 fm. The enhancement of the rms radii
in the excited $1/2^-$ states implies their developed cluster structure.  
The $1/2^-_2$ state at $E_x = 13.8$ MeV largely overlaps
with wave function having $^9\mathrm{Be}+\alpha$
cluster configuration shown in Fig.~\ref{fig:den} (b), which amounts to 0.46. 
Hence, this state has large $S$-factors in the 
$^9\mathrm{Be}(3/2^-_1)\otimes l=2$
and $^9\mathrm{Be}(5/2^-_1)\otimes l=2$ channels that are 0.11 and 0.09, respectively. 
The RWAs in the $^9\mathrm{Be}(3/2^-)\otimes l=2$ and
$^9\mathrm{Be}(5/2^-)\otimes l=2$ channels have two nodes ($N=6$)
as shown in Fig.~\ref{fig:rwa1-} (b), while the those in the ground state have one node ($N=4$). This means that the
$1/2^-_2$ state is regarded as the nodal excitation
of the inter-cluster motion between
$^9\mathrm{Be}$ and $\alpha$ clusters.
Therefore, the $1/2^-_2$ state is not a Hoyle-analogue state but an excited $^{9}{\rm Be}+\alpha$
cluster state,
although it also has non-negligible $S$-factors in the $^{12}\mathrm{C}+n$ channels.

The $1/2^-_3$, $1/2^-_4$ and $1/2^-_5$ states
have large overlap with the basis wave functions displayed in Fig.~\ref{fig:den} (c), (d) and (e),
respectively.
These states have non-negligible $S$-factors in the $^{12}\mathrm{C}(0^+_2)\otimes p_{1/2}$ channel
but the magnitudes are less than 0.20. This means that the $^{12}\mathrm{C}(0^+_2)\otimes p_{1/2}$ 
configuration (Hoyle-analogue configuration) does not manifest as a single excited state,
but it is fragmented into these $1/2^-$ states. 
Thus, we conclude that there is no Hoyle-analogue $1/2^-$ state. 
Interestingly, these states also have the
$S$-factors in the $^{12}\mathrm{C}(2^+_2)\otimes p_{1/2}$ channel, which corresponds 
to the rotational excited state of the Hoyle state.
Difference between the $1/2^-_3$,
$1/2^-_4$ and $1/2^-_5$ states is seen in different magnitudes of the coupling to
the $^9\mathrm{Be}+\alpha$ channels.
Similar result was also obtained by
T. Yamada {\it et al.} \cite{YamadaPRC92}. They argued that Hoyle-analogue state does not appear
in the $1/2^-$ states because of the enhanced $^9\mathrm{Be}+\alpha$ correlation induced by the
attractive odd-parity $\alpha-n$ interaction.  We also confirm this as the non-negligible
$S$-factors in the $^{12}\mathrm{C}(0^+_2)\otimes p_{1/2}$, $^{12}\mathrm{C}(2^+_2)\otimes
p_{3/2}$ and $^{9}\mathrm{Be}+\alpha$ channels. In addition, our result shows the shrinkage of the
rms radii compared to the Hoyle state (2.94 fm), which is also consistent to the interpretation
suggested by T.Yamada {\it et al.} \cite{YamadaPRC92}.

Although there is no Hoyle-analogue state, it is interesting to note that all of four $1/2^-$
states have large monopole transition matrix comparable with the Hoyle state, which, in total,
exhaust 24\% of the energy weighted sum rule which is consistent with the experiment
\cite{Kawa1,Kawa2,Kawa3}. One may wonder why number of the 
excited state with large monopole matrix is increased in $^{13}$C
than in $^{12}$C despite of the fragmentation of the
$^{12}\mathrm{C}(0^+_2)\otimes 0p_{1/2}$ configuration into many
states.  The reason of the increase and the origin of the monopole
strength of each state are explained as follows.

The origin of the monopole strength of the $1/2^-_2$ state is the
excitation of the relative motion between $^9$Be and $\alpha$
clusters.  As already mentioned, the $1/2^-_2$ state has a
$^9\mathrm{Be}+\alpha$ cluster structure in which the inter-cluster
motion is excited by $2\hbar\omega$ from the ground state. Therefore,
it naturally has the enhanced monopole strength.  Different from the
$1/2^-_2$ state, the monopole strengths of the other $1/2^-$ states
originates in the excitation of the $^{12}$C core. In particular,
we found that the monopole excitation of $^{12}\mathrm{C}(2^+_1)
\rightarrow ^{12}\mathrm{C}(2^+_2)$ plays an important role as well
as the excitation of $^{12}\mathrm{C}(0^+_1) \rightarrow
^{12}\mathrm{C}(0^+_2)$.  To elucidate this, we here show a simple
estimation of the monopole transition matrix. First, let us assume
that the ground state of $^{13}$C (the $1/2^-_1$ state) has a
$(0s_{1/2})^4(0p_{3/2})^8(0p_{1/2})^1$ configuration, and
$^{12}\mathrm{C}(0^+_1)$ and $^{12}\mathrm{C}(2^+_1)$ respectively
have a $(0s_{1/2})^4(0p_{3/2})^8$ and
$(0s_{1/2})^4(0p_{3/2})^7(0p_{1/2})^1$ configurations.  Then,
$^{13}\mathrm{C}(1/2^-_1)$ can be written as,
\begin{align}
  \Ket{^{13}\mathrm{C}(1/2^-_1)}
  &= n_{0} \Ket{ \mathcal{A}
  \left\{ {^{12}\mathrm{C}(0^+_1)} \otimes 0p_{1/2} \right\}
  },
  \\
  &= n_{2} \Ket{ \mathcal{A}
  \left\{ {^{12}\mathrm{C}(2^+_1)} \otimes 0p_{3/2} \right\}},
  \label{eq:gswf}
\end{align}
where $n_0$ and $n_2$ denote the normalization factors defined as,
\begin{align}
  n_{0} &= \Braket{\mathcal{A}\left\{{^{12}\mathrm{C}(0^+_1)} \otimes 0p_{1/2}\right\}|\mathcal{A}\left\{ {^{12}\mathrm{C}(0^+_1)} \otimes 0p_{1/2}\right\}}^{-1/2},\\
  n_{2} &= \Braket{\mathcal{A}\left\{{^{12}\mathrm{C}(2^+_1)} \otimes 0p_{3/2}\right\}|\mathcal{A}\left\{ {^{12}\mathrm{C}(2^+_1)} \otimes 0p_{3/2}\right\}}^{-1/2}.
\end{align}
Second, the excited $1/2^-$ states
($1/2^-_3$, $1/2^-_4$ and $1/2^-_5$)
may be written as,
\begin{align}
  &\Ket{^{13}\mathrm{C}(1/2^-_{ex})}
  = a n'_{0} \Ket{ \mathcal{A}
    \left\{ {^{12}\mathrm{C}(0^+_1)} \otimes 0p_{1/2} \right\}
  }
  \nonumber\\
  &+ b n'_{2} \Ket{ \mathcal{A}
\left\{ {^{12}\mathrm{C}(2^+_1)} \otimes 0p_{3/2} \right\}}
  + \text(\rm other\ configurations)
  \label{eq:exwf}
\end{align}
since they are dominated by the
$^{12}\mathrm{C}(0^+_2)\otimes 0p_{1/2}$
and
$^{12}\mathrm{C}(2^+_2)\otimes 0p_{3/2}$
configurations.
Here, $n'_0$ and $n'_2$ are the normalization
factors defined in a similar manner,
and we assumed that the neutron orbits
are unchanged from the ground state. 
We also assume that 
$n_0'\Ket{\mathcal{A}\left\{ {^{12}\mathrm{C}(0^+_1)} \otimes 0p_{1/2} \right\}}$
and
$n_2'\Ket{\mathcal{A} \left\{ {^{12}\mathrm{C}(2^+_1)} \otimes 0p_{3/2} \right\}}$
are orthogonal, and their amplitudes are represented by $a$ and $b$.

Finally, following the discussion
made by T. Yamada {\it et al.} \cite{YamadaPTP120},
we rewrite the monopole 
operator as 
\begin{align}
  \mathcal{M}^{IS0}(^{13}\mathrm{C}) =
  \mathcal{M}^{IS0}(^{12}\mathrm{C}) 
  + \frac{12}{13} r^2,
  \label{eq:is0}
\end{align}
where $\mathcal{M}^{IS0}(^{12}\mathrm{C})$ 
acts on the $^{12}$C core, while $\bm{r}$ 
denotes the 
coordinate between $^{12}\mathrm{C}$ core
and valence neutron.
With these expressions, we can derive
an estimation for the monopole
transition matrix,
\begin{align}
  M(IS0)
  &= \Braket{{^{13}\mathrm{C}(1/2^-_{ex})}|\mathcal{M}^{IS0}(^{13}\mathrm{C})|{^{13}\mathrm{C}(1/2^-_{1})}}
  \nonumber \\
  &= a^*
  \frac{n'_{0}}{n_{0}}
  \Braket{{^{12}\mathrm{C}(0^+_2)}|\mathcal{M}^{IS0}(^{12}\mathrm{C})|{^{12}\mathrm{C}(0^+_1)}}
  \nonumber \\
  &+ b^*
  \frac{n'_{2}}{n_{2}}
  \Braket{{^{12}\mathrm{C}(2^+_2)}|\mathcal{M}^{IS0}(^{12}\mathrm{C})|{^{12}\mathrm{C}(2^+_1)}}
  \nonumber \\
  &+ (other\ channels).
  \label{eq:is0:est}
\end{align}
The derivation of the Eq.~\eqref{eq:is0:est}
is almost same with that explained in Ref.~\cite{YamadaPTP120}.
Thus, the monopole strengths of
the excited $1/2^-$ states
can be related to the monopole transitions
of $^{12}$C.
Here, it is noted that $n'_0/n_0$
and $n'_2/n_2$ are almost equal
to 1, and $\Braket{{^{12}\mathrm{C}(2^+_2)}|\mathcal{M}^{IS0}(^{12}\mathrm{C})|{^{12}\mathrm{C}(2^+_1)}}$
is as large as or even larger than
$\Braket{{^{12}\mathrm{C}(0^+_2)}|\mathcal{M}^{IS0}(^{12}\mathrm{C})|{^{12}\mathrm{C}(0^+_1)}}$.
Therefore, if $a$ and $b$ are not small
and have the same phase, the transition
matrix can be large.
From this simple estimation, it is also
clear that the $^{12}\mathrm{C}(2^+_2)
\otimes 0p_{3/2}$ channel increases
the number of the $1/2^-$
states having large monopole transition
strengths.

\subsection{Structure of the $3/2^-$ states}
\begin{figure}
  \includegraphics[width=0.7\hsize]{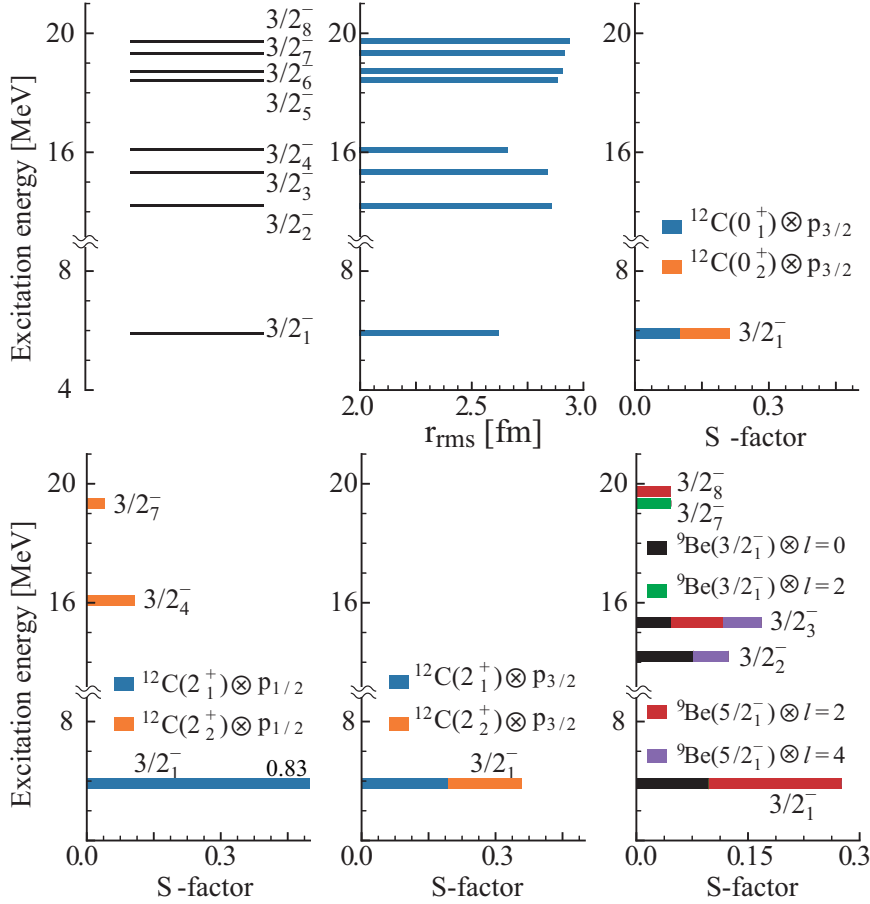}
  \caption{(Color online) The calculated excitation spectra, the rms matter radii $r_{rms}$  and the
    $S$-factors of $3/2^-$ states below 20 MeV. The $S$-factors smaller than 0.05 are not
 displayed.} 
 \label{fig:sfactor3-}
\end{figure}

The $3/2^-$ states are also candidates of Hoyle-analogue state with $P$-wave
valence neutron. The properties of the $3/2^-$ states
below 20 MeV are summarized in Fig.~\ref{fig:sfactor3-}. 
In our calculation, except for the $3/2^-_1$
and $3/2^-_4$ states, the $3/2^-$ states have larger matter rms radius than 2.75 fm.

The $3/2^-_1$ state is obviously dominated by the $^{12}\mathrm{C}(2^+_1)\otimes p_{1/2}$ channel 
($S=0.87$), and its configuration is concluded as $(0p_{3/2})^{-1}(0p_{1/2})^2$ because
$^{12}\mathrm{C}(2^+_1)$ is dominated by the $(0p_{3/2})^{-1}(0p_{1/2})^1$ configuration.
The properties of the other $3/2^-$ states are not clear, 
since their $S$-factors are small in all calculated channels ($S\leq 0.14$).
In particular, there are no state having sizable S-factor in the 
$^{12}{\rm C}(0^+_2)\otimes p_{3/2}$ channel except for the $3/2^-_1$ state. 
Therefore, we conclude that there is no Hoyle-analogue $3/2^-$ state below 20 MeV.

As explained above, the Hoyle-analogue state with
$P$-wave neutron does not appear. This is due to the strong
attractive interaction between $\alpha$ cluster and $P$-wave neutron, which
induces the coupling with many different channels. As a result, the
$^{12}\mathrm{C}(0^+_2) \otimes p_{1/2}$ and $^{12}\mathrm{C}(0^+_2) \otimes
p_{3/2}$ configurations are fragmented into many states.

%\begin{figure}
%  \includegraphics[width=0.8\hsize,clip]{fig/RWA_3-.pdf}
%  \caption{(Color online) The calculated RWAs of the $3/2^-_1$ state in the $^{12}\mathrm{C}+n$ 
%    and $^9\mathrm{Be}+\alpha$ channels. The RWAs which yields the larger $S$ than
%    0.15 are displayed.}
% \label{fig:rwa3-}
%\end{figure}

\subsection{Structure of the $1/2^+$ states}
\begin{figure}
  \includegraphics[width=0.7\hsize]{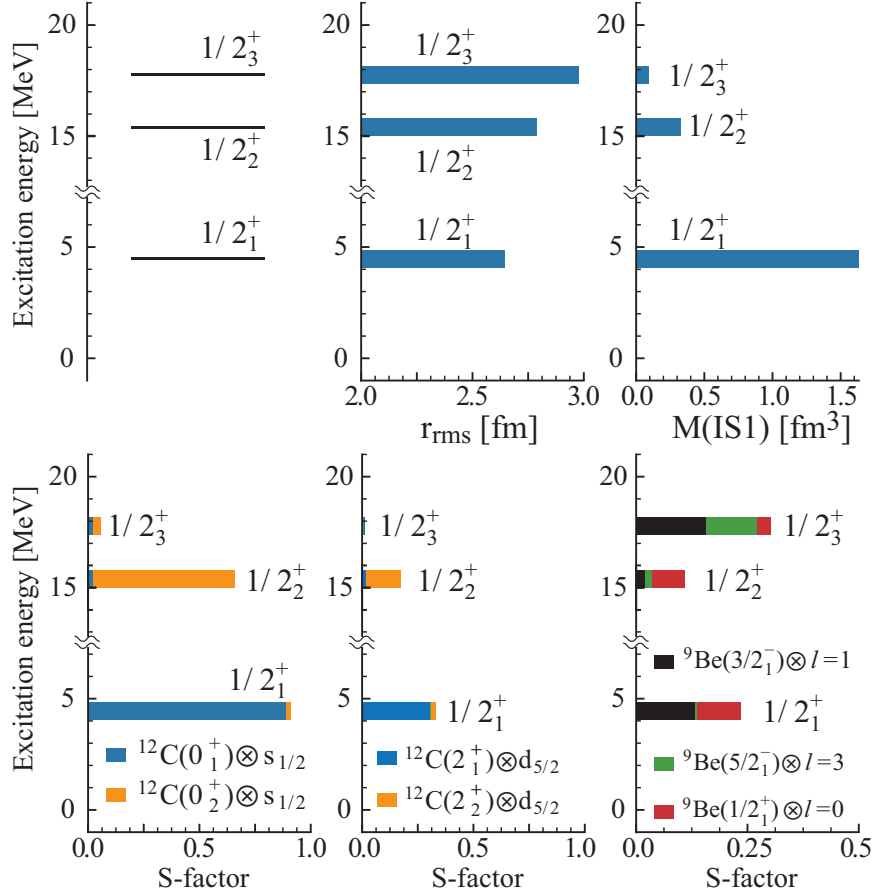}
  \caption{(Color online) The calculated excitation spectra, matter rms radii, $M(IS1)$ and
 $S$-factors of $1/2^+$ states below 20 MeV are shown in a same manner with the
 Fig.~\ref{fig:sfactor1-}.}  \label{fig:sfactor1+}
\end{figure}
In the above discussion, we showed
that there is no Hoyle-analogue state
in the $1/2^-$ and $3/2^-$ states, where
the strong attractive interaction between $\alpha$-clusters
and neutron induces the coupling with many different channels.
On other hand, because the
interaction between $\alpha$ and $S$-wave
neutron is weaker than that for $P$-wave
neutron, we expect that the $1/2^+$ state
is promising candidate of the Hoyle-analogue state.

\begin{figure}
  \includegraphics[width=0.5\hsize]{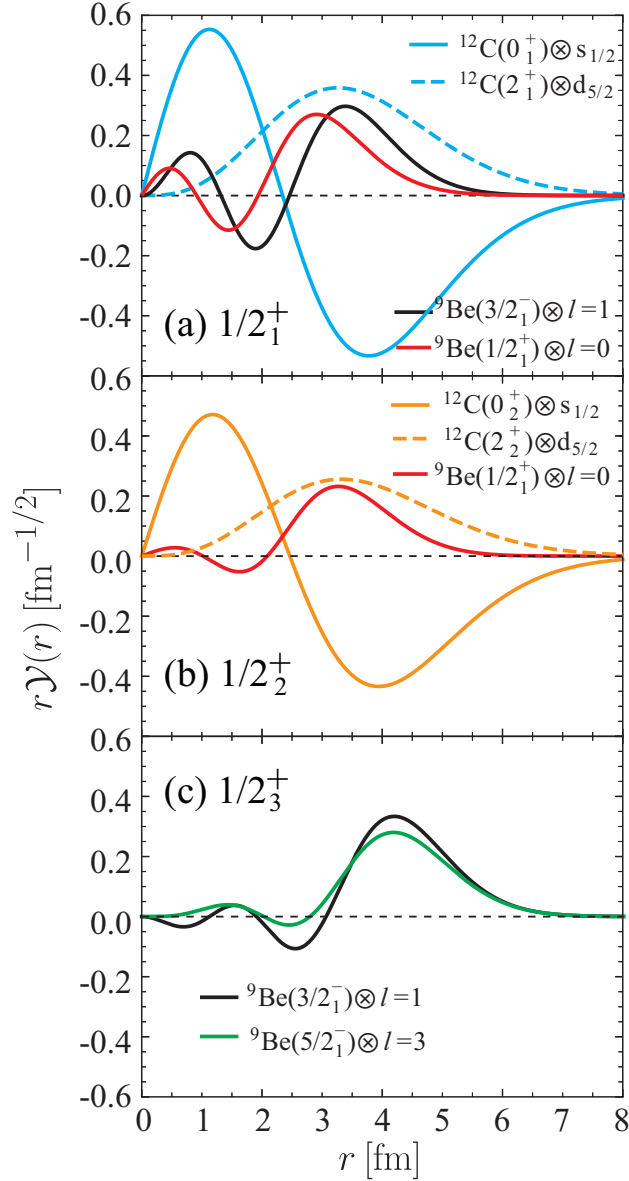}
  \caption{(Color online) The calculated RWAs of the $1/2^+$ states in the $^{12}\mathrm{C}+n$ 
    and $^9\mathrm{Be}+\alpha$ channels. The RWAs which yields the larger $S$ than
    0.04 are displayed.}
 \label{fig:rwa1+}
\end{figure}

The calculated rms radii and $S$-factors are shown in Fig.~\ref{fig:sfactor1+}.
The rms radius of the
$1/2^+_1$ state is only 2.62 fm and suggests
that $1/2^+_1$ state has a
compact shell structure. 
On the other hand, the radii of the $1/2^+_2$ and $1/2^+_3$ states
are larger than 2.75 fm, which 
indicates their developed cluster structure. 
This point can be clearly confirmed by analysis of
the $S$-factor.
The $1/2^+_1$ state is a particle-hole excited state,
because its $S$-factor in the $^{12}\mathrm{C}(0^+_1)\otimes
s_{1/2}$ channel is 0.84 and the other channel contributions are
relatively small. The RWA of the $1/2^+_1$ state in the
$^{12}\mathrm{C}(0^+_1)\otimes s_{1/2}$ has one node (Fig.~\ref{fig:rwa1+}~(a)), and hence, its
particle-hole configuration is $(0p_{1/2})^{-1}(1s_{1/2})^1$.

The $1/2^+_2$ state is located 3.0 MeV above $^{12}\mathrm{C}(0^+_2)+n$ 
threshold energy  and has the largest $S$-factor of 0.64 in the
$^{12}\mathrm{C}(0^+_2)\otimes s_{1/2}$ channel among the calculated $1/2^+$ states.
The RWA in the $^{12}\mathrm{C}(0^+_2)\otimes
s_{1/2}$ channel has one node (Fig.~\ref{fig:rwa1+}~(b)), which indicates that
the $1/2^+_2$ state is the
Hoyle-analogue $1/2^+$ state
with $^{12}\mathrm{C}(0^+_2)\otimes 1s_{1/2}$ configuration.
It has the largest overlap with the basis
wave function shown in the Fig.~\ref{fig:den} (g) but
its magnitude is 0.48. This state also has 
non-negligible overlaps with the various basis wave functions
having $3\alpha+n$  cluster structure, which suggests the dilute gas-like nature of the
$1/2^+_2$ state.  
However, the rms radius of the $1/2^+_2$ state (2.76 fm) is 
shrank compared to that of the Hoyle state (2.94 fm) in our
calculation.
This shrinkage indicates that 
the Hoyle-analogue nature
is weakened by the interaction between $^{12}\mathrm{C}$ and 
valence neutron. 
This point
can be seen in the coupling with the other
$^{12}\mathrm{C}+n$ channels.  For example, the $1/2^+_2$ state has
the non-negligible $S$-factors of 0.11 and 0.18 in the $^{12}\mathrm{C}(1^-_1)\otimes
p_{3/2}$ and $^{12}\mathrm{C}(2^+_2)\otimes d_{5/2}$ channels, respectively.

The $1/2^+_3$ state has quite different nature from the $1/2^+_1$ and $1/2^+_2$
states. This state has almost zero $S$-factors 
in the $^{12}\mathrm{C}+n$ channels,
but the $S$-factors in the $^{9}\mathrm{Be}+\alpha$ channels
amounts to 0.32 in total.
Therefore, we conclude
that the $1/2^+_3$ state has the $^{9}\mathrm{Be}+\alpha$ cluster structure.
Interestingly, its density distribution (Fig.~\ref{fig:den} (h)) shows
the similar structure to the $1/2^-_2$ state (Fig.~\ref{fig:den} (b)).
Furthermore, the $S$-factors indicate that 
they are dominated by the $^9\mathrm{Be}(3/2^-_1)+\alpha$
and $^9\mathrm{Be}(5/2^-_1)+\alpha$ channels.
Therefore, we consider that the $1/2^-_2$ and $1/2^+_3$ state constitute the parity-doublet having
the $^9\mathrm{Be}+\alpha$ cluster structure.
A similar bent-armed $3\alpha+n$
cluster structure in negative parity states
was also discussed by N.~Furutachi {\it et al.} \cite{Furutachi} in relation with 
the inversion-doublet of $^9\mathrm{Be}+\alpha$ cluster band suggested by M.~Millin and W. von
Oertzen \cite{Millin}.  The predicted $3/2^-_2$ state has the similar bent-armed $3\alpha+n$
cluster structure to the $1/2^-_2$ and $1/2^+_3$ states in the present calculation. 

The calculated IS1 transition matrix $M(IS1)$ (Fig.~\ref{fig:sfactor1+})
indicate that the $1/2^+_1$ state is strongly populated by IS1 transition
from the ground state ($M(IS1) = 0.95$ W.u.) 
while the $1/2^+_2$ and $1/2^+_3$ states are
not. In particular, despite of its Hoyle-analog
structure, the IS1 transition strength
of the $1/2^+_2$ state is unexpectedly small.
This may be explained as follows. Following
the discussion in Refs.~\cite{ChibaPRC93,KimuraPRC95}, we decompose
the system into $^{12}$C core
and the valence neutron, and rewrite the
IS1 operator as
\begin{align}
  &\mathcal{M}^{IS1}_{\mu} = 
  \frac{132}{169} r^2\mathcal{Y}_{1\mu}(\bm{r})
  - \frac{5}{39}\sum_{i\in ^{12}\mathrm{C}}\xi^2_i 
      \mathcal{Y}_{1\mu}(\bm{r})
      \nonumber \\
      &+ \frac{4\sqrt{2\pi}}{39}\left[\sum_{i\in ^{12}\mathrm{C}}\mathcal{Y}_{2}(\bm{\xi}_i) \otimes
      \mathcal{Y}_{1}(\bm{r})\right]_{1\mu}
      + \sum_{i\in {^{12}\mathrm{C}}} \xi^2_i \mathcal{Y}_{1\mu}(\bm{\xi}_i)
  \label{eq:IS1:Cn}
\end{align}
where $\bm{\xi}_i$ denote the internal coordinates of the $^{12}$C core, while $\bm{r}$ denotes
the valence neutron coordinate.  The first term of the Eq.~\eqref{eq:IS1:Cn}
is dependent only on $\bm{r}$ and induces the IS1 transition of the valence neutron. Therefore, 
the $1/2^+_1$ state which has the $1p1h$ configuration
is mainly excited by this term. This is the reason why $1/2^+_1$ state 
has strong IS1 transition matrix comparable with Weisskopf estimate.

On the other hand,
the second and third terms induce
the monopole and quadrupole transitions
of $^{12}$C core. Therefore,
if these terms act on the ground state wave
function (Eq.~\eqref{eq:IS1:Cn}), they bring about the
core excitations 
$^{12}\mathrm{C}(0^+_1) \rightarrow 
^{12}\mathrm{C}(0^+_2)$
and
$^{12}\mathrm{C}(2^+_1) \rightarrow 
^{12}\mathrm{C}(0^+_2)$
combined with the valence neutron excitations
to yield the Hoyle-analogue $1/2^+_2$ state.
Since the transition matrix for the
core excitations are large,
we expect that the IS1 excitation from the
ground state to
the $1/2^+_2$ state is
enhanced. However, the coefficients
for these two terms are rather small
($5/39$ and $4\sqrt{2\pi}/39$).
As a result, the $1/2^+_2$ state
has relatively small transition strength
despite of its dilute gas-like nature.

Thus, the Hoyle-analogue $1/2^+_2$ state has unexpectedly small $M(IS1)$. However,
in the preliminary reported IS1 transition strength distribution of
$^{13}\mathrm{C}$,
there is a small peak around 13 MeV \cite{Kawa1,Kawa2,Kawa3}, which is close 
to the our prediction
and may correspond to the Hoyle-analogue
$1/2^+_2$ state.

\begin{figure}
  \includegraphics[width=0.8\hsize]{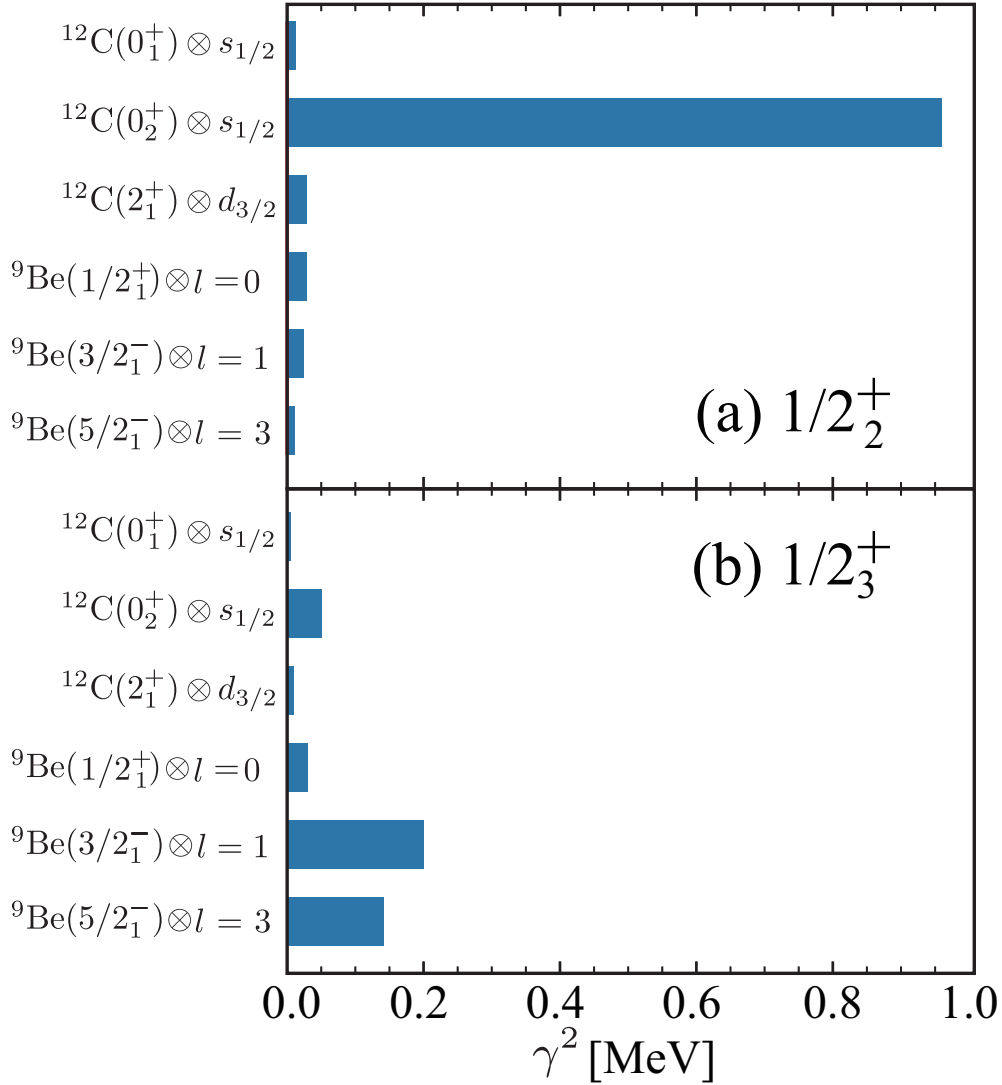}
  \caption{(Color online) The calculated reduced widths $\gamma^2$ of the $1/2^+_2$ and $1/2^+_3$ states 
    for $^{12}\mathrm{C}+n$ and $^9\mathrm{Be}+\alpha$ channels. The matching radius $a= 4.5$
    fm is applied.}
 \label{fig:rw1+}
\end{figure}

Finally, we  discuss the decay width
of the $1/2^+$ states. Owing to the cluster structure of the $1/2^+_2$ and $1/2^+_3$ states, 
they have unique decay patterns. The
calculated the reduced width $\gamma^2$ of the $1/2^+_2$ and $1/2^+_3$
states in the $^{12}\mathrm{C}+n$
and $^9\mathrm{Be}+\alpha$ channels are shown in Fig.~\ref{fig:rw1+}. 
The $1/2^+_2$ state has largest reduced width of 0.96 MeV in the $^{12}\mathrm{C}(0^+_2)\otimes
s_{1/2}$ channel. The reduced widths 
in the other channels are negligibly small. On the other
hand, The reduced decay widths in the $^9\mathrm{Be}(3/2^-_1) \otimes l=1$
and $^9\mathrm{Be}(5/2^-_1) \otimes l=3$ channels are largest in the  
$1/2^+_3$ state.
Because of the larger $Q$-value, the $1/2^+_3$ state
dominantly decays via the $^9\mathrm{Be}(3/2^-_1) \otimes l=1$
channel.
Therefore, the strong decays via
the $^{12}\mathrm{C}(0^+_2)\otimes s_{1/2}$ and
$^9\mathrm{Be}(3/2^-_1) \otimes l=1$ channels
are signature of the $1/2^+_2$ and $1/2^+_3$ states,
respectively.

\section{Summary}
We studied the Hoyle-analogue states in
$^{13}$C based on AMD.
The basis wave functions are obtained by 
the energy variation with constraint on the expectation values of harmonic
oscillator quanta. Using these basis
wave functions,   GCM calculation
was performed to obtain
the excitation energies and the eigen wave functions.

The analysis of the $S$-factors in the $^{12}\mathrm{C}+n$ and $^9\mathrm{Be}+\alpha$
channels revealed the characters of the ground and excited states of
$^{13}\mathrm{C}$.  The ground state ($1/2^-_1$ state) has the
$(0s)^4(0p_{3/2})^8(0p_{1/2})^1$ configuration, and the $3/2^-_1$ and $1/2^+_1$
states have the $1p1h$ configurations.
In contrast to these
shell model like states, the non-yrast states have developed cluster structure. The $1/2^-_2$
and $1/2^+_3$ states constitute the inversion-doublet of the
bent-armed $^9\mathrm{Be}+\alpha$ cluster structure. The $1/2^-_3$, $1/2^-_4$ and
$1/2^-_5$ states are $3\alpha+n$ cluster states in which the
$^{12}\mathrm{C}(0^+_2)\otimes 0p_{1/2}$ and $^{12}\mathrm{C}(2^+_2)\otimes 0p_{3/2}$
configurations are mixed. However, they cannot be regarded as the Hoyle-analogue state because the  
$S$-factors in the $^{12}\mathrm{C}(0^+_2)\otimes p_{1/2}$ channel are small.
Similarly, there is no Hoyle-analogue state
in $3/2^-$ states, because of the
fragmentation of $^{12}\mathrm{C}(0^+_2)\otimes 0p_{3/2}$ configuration
into many states. The absence of the
Hoyle-analogue states in $P$-wave states
is attributed to the strong $\alpha -n$ $P$-wave
interaction.
On the hand, the $1/2^+_2$ state located at 15.4 MeV is
Hoyle-analogue state dominated by the $^{12}\mathrm{C}(0^+_2)\otimes 1s_{1/2}$
configuration with $S=0.64$.

The characters of the $1/2^-$ and $1/2^+$ states are reflected to the IS0 and
IS1 transitions. The IS0 transitions to the excited $1/2^-$ states are
comparable to the Hoyle state in $^{12}$C.
The origins of the enhanced IS0 transitions are clustering nature of the excited
$1/2^-$ states. In particular, the enhanced $M(IS0)$ of the $1/2^-_3$,
$1/2^-_4$ and $1/2^-_5$ originate in the coupling of
the $^{12}\mathrm{C}(0^+_2)\otimes 1s_{1/2}$ and $^{12}\mathrm{C}(2^+_2)\otimes 0p_{3/2}$ configurations.
Contrary, the IS1 transition to the Hoyle-analogue $1/2^+_2$ state
is suppressed due to the property of the IS1 transition operator.

The decay widths of the $1/2^+$ states show very unique patterns. The 
Hoyle-analogue $1/2^+_2$ state dominantly decays via the $^{12}\mathrm{C}(0^+_2)\otimes s_{1/2}$
channel but the $1/2^+_3$ state decays via $^{9}\mathrm{Be}(3/2^-_1)\otimes l=1$ channel.
These unique decay patterns are key observable to identify the Hoyle-analogue $1/2^+$ state.

\begin{acknowledgements}
The authors acknowledge that the discussion with Dr. Kawabata and Dr. Taniguchi was fruitful for
this work. This work was supported by JSPS KAKENHI Grant No. 16K05339. 
\end{acknowledgements}

\end{document}